# Flexible Quasi-Yagi-Uda antenna for 5G communication

Behzad Ashrafi Nia, Franco De Flaviis
Department of Electrical Engineering and Computer Science
University of California, Irvine
Irvine, USA
bashrafi@uci.edu, franco@uci.edu

Soheil Saadat
Multi-Fineline Electronix Inc. (MFLEX)
Irvine, USA

soheil2054@gmail.com

*Abstract*— This paper presents the design and experimental of a single and array Quasi Yagi-Uda antenna at 28 GHz. The proposed antenna is implemented on MFLEX flexible material with a thickness of 0.120mm for 5G applications. A wideband antenna operates within 24 to 29.5 GHz and exhibits almost the same end-fire radiation pattern over bandwidth with an average gain of 6.2dBi and 10.15dBi for single and array antennas. The flexible antenna was tested under bending conditions and results showed excellent performance at the 28GHz region.

## I. Introduction

The Fifth-generation (5G) mobile communication is considered as very attractive solutions for exponentially expanding wireless data traffic in the future, due to its ability to use of significantly larger bandwidths at millimeter-wave (mm-wave) frequencies which have been available by the U.S. Federal Communications Commission (FCC). This frequency range including a new licensed spectrum located in the 28 GHz (27.5–28.35 GHz), 37 GHz (37–38.6 GHz), and 39 GHz (38.6–40 GHz) bands. Unfortunately, switching to the mm-Waves would bring us new challenges and certainly needs more consideration on the antenna design for 5G devices. One of the limitations with the mm-Waves wireless communication is the increased path loss because of large propagation attenuation at mm-Wave frequencies. In order to overcome this obstacle, the antenna array with wideband, high gain, and the ability to generate directional beams, stable radiation characteristics are desirable. The small wavelengths of mm-Wave frequencies facilitate the use of multiple number antenna elements in a compact size and enable us to have higher gain. Among different type of the antennas, such as patch, slot, the end-fire antennas including Quasi-Yagi, Dipole and Vivaldi are more suitable to cover the required cover-space of 5G cellular communication [1, 2, 3]. Among them, the Quasi-Yagi antenna has more potential due to its high gain, easy fabrication, stable radiation pattern, and compact size. In this study, we have selected the frequency band of 28GHz as the design target. Design of compact Quasi-Yagi–Uda antennas with end-fire radiation are presented for mm-Wave wireless applications. The proposed Quasi-Yagi antenna is on a low-cost, single-substrate-layer flexible printed circuit (FPC). Flexible substrate materials have brought advanced electronic circuits into products ranging from smartphones to medical devices [4, 5]. The ability of flex circuits to accommodate tight bending radii and eliminate the need for cables and connectors gives designers greater flexibility and enables products that would not otherwise be possible. Another key factor for flex material is to have low loss tangent across a wide range of frequencies, including 28 and 38 GHz, which make them a good choice for 5G mm-wave applications.

## II. Quasi-Yagi Antenna design

The schematic of the proposed antenna is shown in Fig. 1. which has GCPW feeding. The antenna is constructed on the modified polyimide (MPI) material by MFLEX, which is designed for high-speed flex applications. Key property advantages of this material are Low dielectric constant (2.85), Low loss tangent (0.004), Tight thickness tolerance, Better bendability, and Low moisture absorption. The total thickness of the substrate in this design is only 120um. This antenna is fed by a grounded coplanar waveguide (GCPW) to a microstrip line and a dipole, which is designed by half-wavelength resonance at 28GHz. A microstrip to parallel-strip balun is used to feed the dipole in balanced mode. Then two arms of each dipole are printed on both sides of the substrate. Also, two directors are employed in front of the dipole, which directs the antenna propagation toward the end-fire direction and helps to improve the directivity of the radiation pattern. Moreover, these elements act as an impedance matching element. Meanwhile, two sets of the reflectors are used on both sides of the substrate for effective reflection of back-radiated electromagnetic waves.

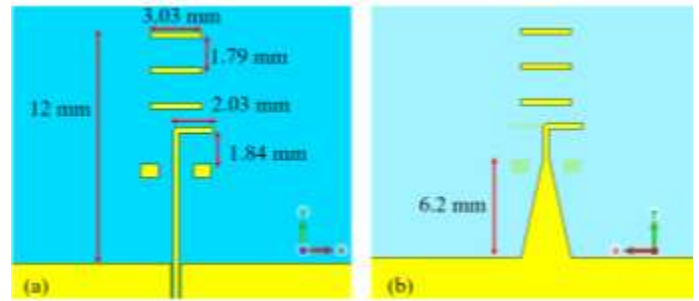

Fig. 1. Structure of proposed antenna. (a) top viewe and (b) back view.

After successfully developed and design of a single element of the Quasi-Yagi antenna, this study also explores the schematic and characterization of the array antenna with four elements of the Quasi-Yagi antenna for applications in 5G platforms. As you can see in Fig.2, four elements of the antenna are placed on both sides of the substrate and fed by T junction power divider.

III. ANTENNA CHARACTRERIZATION AND MEASUREMENTS

Fabricated Quasi-Yagi antenna is shown in Fig. 2. The full-wave simulation was performed using CST STUDIO SUITE. All the measurements were done by a far-field probe station system, and the GSG-250-P probe was used to excite the GCPW line. Simulated and measured S-parameters are shown in Fig. 2, where good agreement is obtained. Both single and array antenna have a measured bandwidth from 23 to 30GHz; thus, a wide frequency coverage is realized. Measured pick gain also is shown for both single element and array antenna in Fig. 3. The measured gain is almost stable in the range of the operation and is in the range of 5.2~6.2 dBi for a single element and 9.2~10.2 dBi for an array antenna.

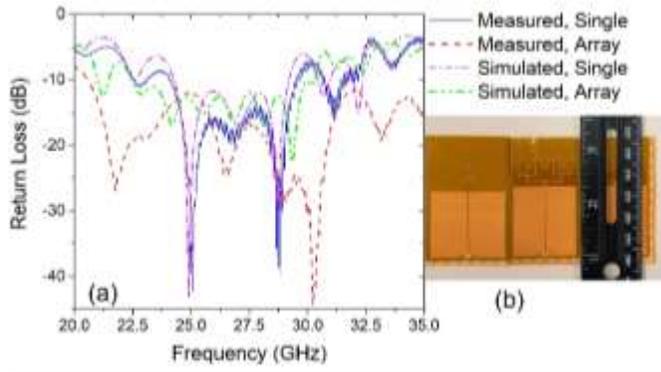

Fig. 2. (a) Measured and CST simulation of S parameters. (b) fabricated antennas.

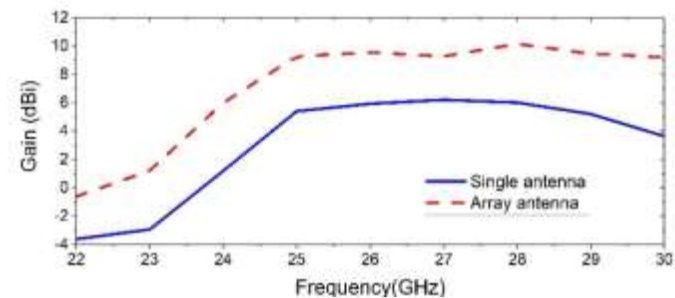

Fig. 3. Measured Gain of single and array Quasi-Yagi-Uda antenna over frequency.

Measured and simulated far-field radiation pattern at 28GHz are compared in Fig. 4. It can be observed the measurement results approved the simulation in both E-plane and H-plane. The patterns of the single antenna show the HPBW of 50° and 58° in the H-and E-planes, and array antenna show the HPBW of 60° and 21° in the H-and E-planes, respectively.

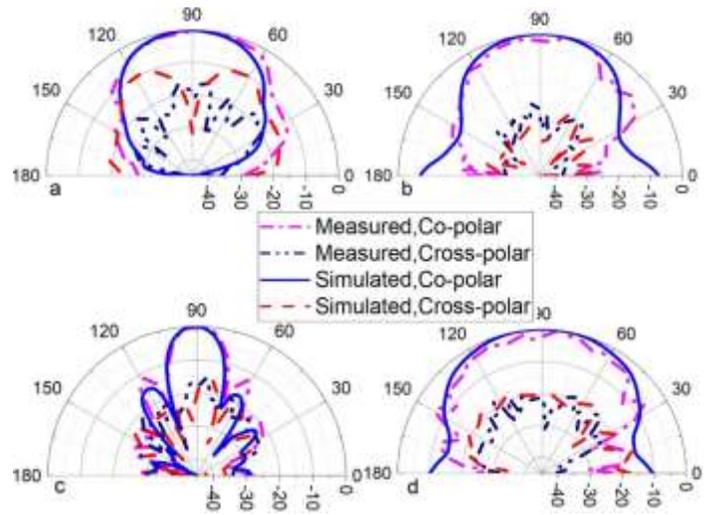

Fig. 4. Measured and CST simulation radiation pattern at 28GHz. (a) At E-plane (single). (b) At H-plane (single). (c) At E-plane (Array). (c) At H-plane (Array).

IV. CONCLUSION AND FUTURE WORK

In this paper, a high gain, wideband Yagi-Uda antenna at 28GHz with the vertical end-fire radiation has been proposed. The antenna has been designed on a flex substrate. The design has resulted in the good agreements between the simulated and measured in the single and array. Measurements show the average realized gain of the single antenna is about 6.2dBi, impedance bandwidth of 25.7% covering from 23 to 29.8 GHz, and the radiation efficiency is up to 88%. Therefore, the proposed antenna is a good candidate for the 5G commination systems.

ACKNOWLEDGMENT

The authors acknowledge the support of Dassault Systèmes Simulia Corp for providing the simulation software (CST STUDIO SUITE). Also, this project was funded and manufactured by Multi-Fineline Electronix Inc. (MFLEX).